\documentclass[12pt]{iopart}

\usepackage{epsfig}
\usepackage{diplom}

\begin{document}
\title{Search for direct photons in p+Pb and p+C collisions at $\sqrt{s_\mathrm{NN}} = 17.4$~GeV}
\author{C Baumann for the WA98 collaboration}
\address{University of M\"{u}nster, Institut f\"{u}r Kernphysik, Wilhelm-Klemm-Str. 9, 48149 M\"{u}nster, Germany}
\ead{cbaumann@uni-muenster.de}

\begin{abstract}
Upper limits on direct photon production  were determined as a function of the transverse momentum for $0.7 < p_\mathrm{T} \leq 3.2$ GeV$/c$
with the WA98 experiment in p+C and p+Pb collisions at $\sqrt{s_\mathrm{NN}} = 17.4 \mathrm{~GeV}$.
The results are compared to direct photon measurements in Pb+Pb collisions at $\sqrt{s_\mathrm{NN}} = 17.3 \mathrm{~GeV}$ by WA98. Implications for a possible thermal direct photon contribution are discussed.
\end{abstract}

\submitto{\JPG}

\section{Introduction}

Direct photons are a unique probe of the early stage of a heavy-ion collision, as they leave the interaction region
without being influenced by strong interactions or hadronization processes. They are therefore regarded as an important tool in the search
for the quark-gluon plasma in ultrarelativistic heavy-ion collisions.
As direct photons from different sources, e.g. prompt and thermal direct photons, cannot be separated experimentally,
further information is needed to quantify the different contributions.

The WA98 collaboration was the first to publish a significant direct-photon excess in heavy-ion collisions at $\sqrt{s_\mathrm{NN}} = 17.3 \mathrm{~GeV}$~\cite{GamPRLWA98}. The excess could be understood by theoretical descriptions with and without a thermal direct photon contribution~\cite{TurRapGal}. To pin down a thermal component, the Pb+Pb data were compared to p+p and p+A data at $\sqrt{s} = 19.4 \mathrm{~GeV}$~\cite{GamPRLWA98}. Due to large systematic uncertainties introduced by the necessary $x_\mathrm{T}$-scaling and a limited $p_\mathrm{T}$ coverage in the region where a thermal contribution would be most prominent, no limits on the prompt photon contribution could be derived~\cite{GamLongWA98}. pQCD calculations have large uncertainties at $\sqrt{s} \approx 20\mathrm{~GeV}$~\cite{Aurenche}, therefore no definite conclusions on a thermal contribution could be drawn based on the available information~\cite{Stankus,Reygers}.

Here, we will present results on direct photon production in p+Pb and p+C collisions measured at $\sqrt{s_\mathrm{NN}} = 17.4 \mathrm{~GeV}$ with the WA98 experiment in 1996. These results will be compared to the Pb+Pb data.

\section{Experimental Results}
In the WA98 experiment photons were measured using a highly segmented lead glass calorimeter, positioned 21.5 m downstream of the target and covering the pseudorapidity range $2.3 \leq \eta \leq 3.0$. The minimum bias trigger was determined by measuring the transverse energy ($E_\mathrm{T}$) with a hadronic calorimeter in the range $3.5 \leq \eta \leq 5.5$.
The measured minimum bias cross section $\sigma_{mb}$ for p+C (p+Pb) of 170 mb (1341 mb) corresponds to $74\%$ ($76\%$) of the total geometric cross section. To enhance the reach in $p_{\mathrm{T}}$, a high-energy photon (HEP) trigger based on the energy signal in the lead glass calorimeter was used.

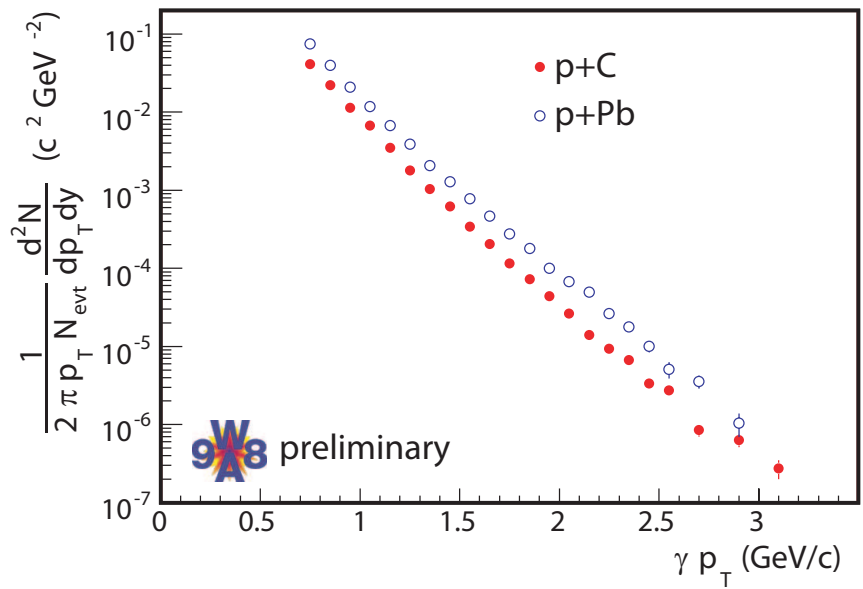
\EPS{pAIncGamma}{width=0.81\textwidth}{Invariant inclusive photon yields in minimum bias p+C and p+Pb collisions. The error bars represent the quadratic sum of systematic and statistical uncertainties.}

The fully corrected inclusive photon yields in p+C and p+Pb collisions at $\sqrt{s_\mathrm{NN}} = 17.4 \mathrm{~GeV}$ are presented in Figure \ref{pAIncGamma}. The transition from the minimum bias to the HEP data sample occurs at $p_\mathrm{T} = 1.2 \mathrm{~GeV}/c$.

In order to get the direct-photon excess $\gamma_\mathrm{direct}$, the contribution $\gamma_\mathrm{decay}$ from decay photons has to be  subtracted from the inclusive photon yield $\gamma_\mathrm{inclusive}$. This can be done by calculating the direct-photon yield $\gamma_\mathrm{direct}$ as a fraction of the inclusive photon spectrum:
\begin{equation}
\gamma_\mathrm{direct} = \gamma_\mathrm{inclusive} - \gamma_\mathrm{decay} = \left(1- \frac{1}{R_\gamma} \right) \cdot \gamma_\mathrm{inclusive} ;\quad R_\gamma = \frac {(\gamma / \pi^0)_\mathrm{meas}}{(\gamma / \pi^0)_\mathrm{sim}}
\end{equation}
The neutral pion yields have been presented in \cite{MSWA98}. A Monte-Carlo simulation based on the measured neutral pion yields was used to determine the simulated $\gamma / \pi^0$-ratio. The simulation uses a parameterization of the measured neutral pion spectrum as input and simulates the photonic decays of the relevant mesons, e.g. $\eta, \eta', \omega$.

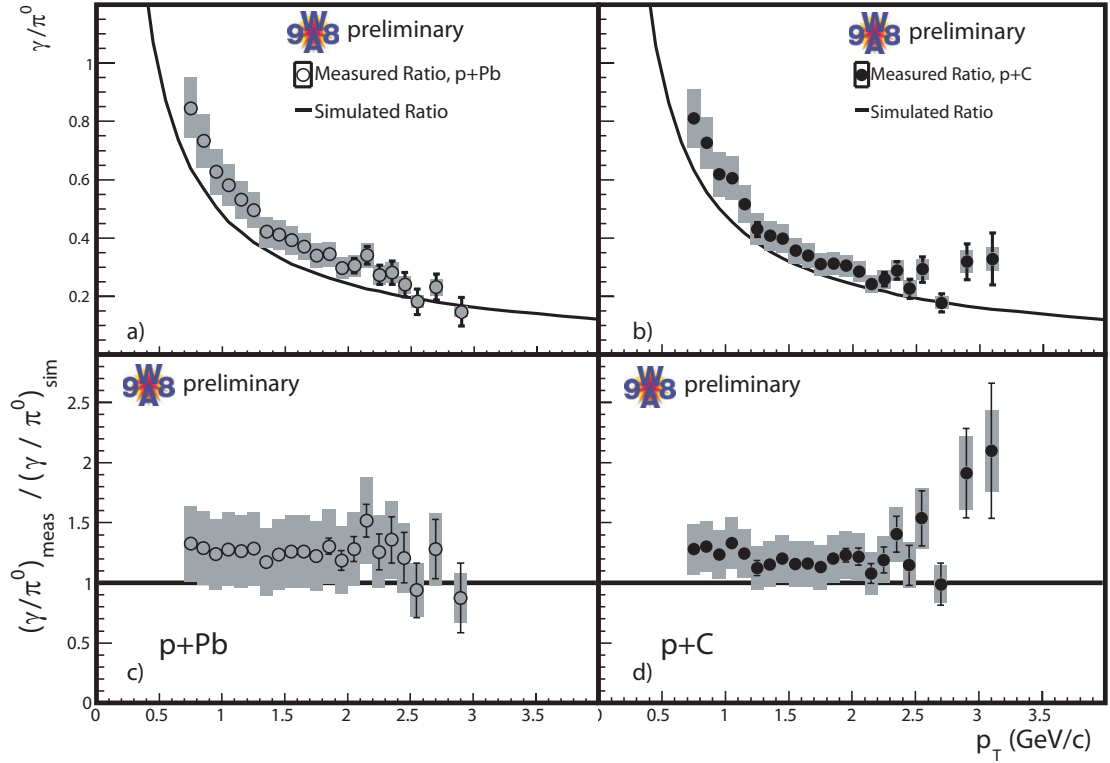
\EPS{Ratios}{width=\textwidth}{ a) Ratio of the inclusive photon yield to the neutral pion yield in p+Pb collisions, the boxes indicate systematic uncertainties, the error bars the statistical uncertainties. The black line indicates the ratio expected from the decay of $\pi^0$, $\eta$ and other hadrons determined with a Monte-Carlo simulation.
    b) The same for p+C collisions.
    c) The double-ratio of the measured to the simulated $\gamma / \pi^0$-ratio for p+Pb collisions. The boxes indicate systematic, the error bars statistical uncertainties. d) The same for p+C collisions.}

The simulated and measured $\gamma / \pi^0$-ratios and the corresponding double-ratio $R_\gamma$ for p+Pb and p+C collisions are presented in Figure \ref{Ratios}. While the measured $\gamma / \pi^0$-ratio is systematically above the simulated one for both targets, $R_\gamma$ is consistent with unity in the $p_\mathrm{T}$ range relevant for a thermal contribution. The resulting direct-photon yields are presented in Figure \ref{pASpectraPanel}. Significant data points can only be quoted for the highest transverse momenta.

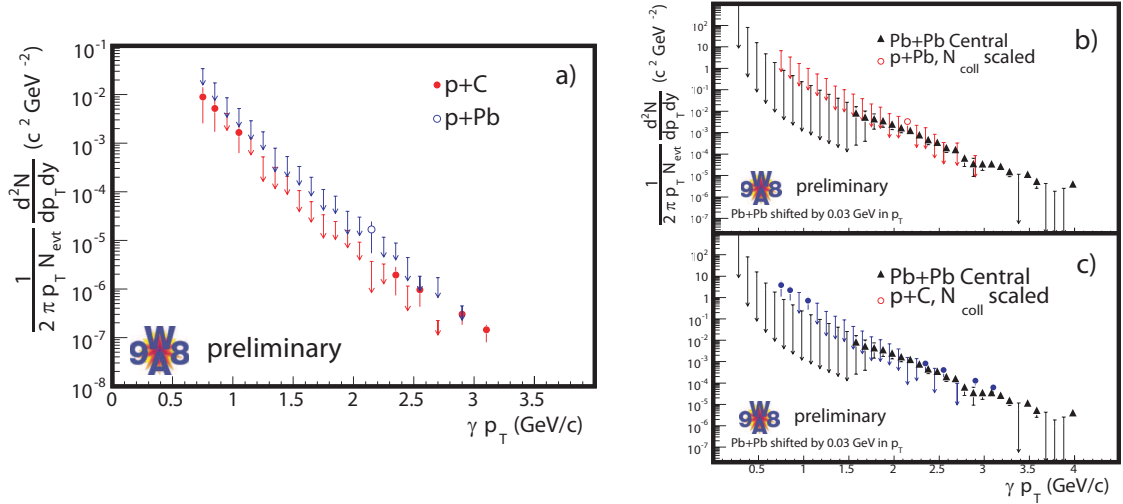
\EPS{pASpectraPanel}{width=\textwidth}{a) Invariant direct photon yields in p+C and p+Pb collisions. The error bars represent the quadratic sum of statistical and systematic uncertainties. The upper edges of the arrows indicate upper limits (best estimate + $1.28 \sigma$). b) Comparison of the direct photon results from p+Pb ($N_\mathrm{coll} = 3.8$) scaled with the number of binary collisions and the central Pb+Pb direct photon spectra ($N_\mathrm{coll} = 727.8$). c) The same for p+C collisions ($N_\mathrm{coll} = 1.7$).}


As the p+A data should not exhibit a thermal contribution, any significant excess of the direct photons from Pb+Pb collisions over the p+A results, scaled by the average number of nucleon-nucleon collisions $N_\mathrm{coll}$ as described in \cite{MSWA98}, can be attributed to a thermal source. The yields are compared in Figure~\ref{pASpectraPanel}. As both data sets agree within errors, no further limit on the contribution of prompt direct photons can be set.


\section{Summary}
Direct photon yields in p+C and p+Pb collisions at $\sqrt{s_\mathrm{NN}} = 17.4 \mathrm{~GeV}$ were measured by the WA98 experiment. Upper limits and invariant photon yields for direct photons as a function of $p_\mathrm{T}$ could be derived from both data sets. When compared to the results on direct photon production in Pb+Pb collisions by applying $N_\mathrm{coll}$ scaling, no additional constraints on the prompt direct photon production in this system can be set from the new p+C and p+Pb data.


\end{document}